\documentclass[final,3p,times,twocolumn]{elsarticle}

\usepackage{amssymb}
\usepackage{amsmath}

\usepackage{lineno}
\usepackage{booktabs}
\usepackage{xcolor}
\usepackage{multirow}
\usepackage{url}
\usepackage{hyperref}
\usepackage{overpic}
\usepackage{upgreek}

\journal{Nuclear Instruments and Methods A}

\begin{document}

\begin{frontmatter}



\title{Calibration of diamond detectors for dosimetry in beam-loss monitoring}

\author[UniTS,SNS]{G. Bassi}
\author[INFN]{L. Bosisio}
\author[INFN]{P. Cristaudo}
\author[INFN]{M. Dorigo\corref{corr}}
\cortext[corr]{corresponding author}
\ead{mirco.dorigo@ts.infn.it}
\author[UniTS,INFN]{A. Gabrielli}
\author[INFN]{Y. Jin}
\author[UniTS,INFN,IPMU]{C. La Licata}
\author[INFN]{L. Lanceri}
\author[UniTS,INFN]{L. Vitale}

\address[UniTS]{Dipartimento di Fisica, Universit\`a di Trieste, I-34127 Trieste, Italy}
\address[INFN]{INFN, Sezione di Trieste, I-34127 Trieste, Italy}
\address[IPMU]{now at: Kavli Institute for the Physics and Mathematics of the Universe (WPI), University of Tokyo, Kashiwa 277-8583, Japan}
\address[SNS]{now at: Scuola Normale Superiore, I-56126 Pisa, Italy}

\begin{abstract}
Artificially-grown diamond crystals have unique properties that make them suitable as solid-state particle detectors and dosimeters in high-radiation environments. We have been using sensors based on single-crystal diamond grown by chemical vapour deposition  for dosimetry and beam-loss monitoring at the SuperKEKB collider. Here we describe the assembly and the suite of test and calibration procedures adopted to characterise the diamond-based detectors of this monitoring system. We report the results obtained on 28  detectors and assess the stability and uniformity of  response of these devices.

\end{abstract}


\end{frontmatter}


\section{Introduction}
\label{sec:introduction}
Natural diamonds, selected for high purity, have been known as good particle detectors with very interesting properties since a rather long time~\cite{ref_natural,CANALI1979}. Practical applications became feasible with the advent of new techniques for the artificial growth of diamond crystals and their steady improvement. Since the initial proposals for applications in high-energy physics experiments~\cite{ref_hep}, the growth process by chemical vapour deposition (CVD)~\cite{ref_cvd} has been progressing towards the production of single-crystals (sCVD) and poly-crystals (pCVD) of larger size, higher purity and reduced variability, at decreasing cost~\cite{ref:cvd_progress}. At present, artificial-diamond applications range from medical micro-dosimetry~\cite{ref_micro_dose1,ref_micro_dose2} to beam-condition and beam-loss monitors for high-energy accelerators and experiments: examples include the BaBar experiment at PEP-II (SLAC), CDF at Tevatron (Fermilab), ATLAS and CMS at LHC (CERN)~\cite{ref_babar,ref_cdf,ref_atlas,ref_cms}.

Diamond is an insulating material  due to the large energy gap of about $5.5$\,eV between the valence and conduction bands. Diamond sensors, in their simplest planar geometry, are electrically polarised by metal electrodes on two opposite sides, and act as solid-state ionisation chambers. 
Drifting towards the electrodes, electron-hole pairs created by impinging radiation induce a current in the polarising circuit. Under a quasi-stationary flux of radiation, the induced current is proportional to the energy released by the radiation per unit time, and it can be used to measure the dose rate. 

We designed, built, and have been operating a system employing 28 diamond-based detectors for dosimetry and beam-loss monitoring in the interaction region of the SuperKEKB electron-positron collider for the Belle~II experiment~\cite{ref_superkekb,ref_belle2}. The main requirements for these detectors are: a wide range of radiation dose-rates, from a few $\mathrm{\upmu rad/s}$ to several hundred $\mathrm{krad/s}$, and performance stability in locations where an integrated dose in excess of 10\,Mrad can be expected through the lifetime of the experiment.  
Reference~\cite{ref:performance} details the performance of this monitoring system during the first two years of SuperKEKB operations.

In this paper, we report on  mechanical assembly, test and calibration procedures of the diamond detectors employed in such a system, aiming at a comparative assessment of their performance for our radiation-monitoring purposes.

\section{Detector assembly and dark currents}
\label{sec:detectors_assembly}
After carrying out some preliminary  
studies to compare the response of pCVD and sCVD diamond sensors equipped with electrodes of different materials,
we opted for sCVD diamonds from Element Six~\cite{ref_e6} with electrodes processed by CIVIDEC~\cite{ref_cividec}.
The specifications of the supplied sensors are:
sCVD diamond with $(4.5 \times 4.5)$\,mm$^{2}$ faces and $0.50$\,mm thickness, with a tolerance of   $+0.2$/$-0$\,mm and  $\pm 0.05$\,mm for the lateral dimensions and the thickness, respectively; 
$(4.0 \times 4.0)$\,mm$^{2}$ electrodes on both faces,  made of  Ti+Pt+Au layers with  $(100$+$120$+$250)$\,nm thickness.  

The sensors passed the following quality controls before delivery by CIVIDEC:
(i)~maximum dark current within $\pm 500$\,pA in the bias-voltage range $\pm 500$\,V, and within $\pm 20$\,pA in the range $\pm 200$\,V;
(ii)~average ionisation-energy less than $25$\,eV and regular pulse shape from a measurement with  $\upalpha$ particles from $^{241}$Am decays; 
 (iii)~stable current 
 through one-hour irradiation with $\upbeta$ electrons from $^{90}$Sr decays, 
with bias voltages of $\pm100$\,V and $\pm200$\,V;
(iv)~visual inspection and rejection in case of deep scratches.

\begin{figure}[t]
	\centering
		\begin{overpic}[width=0.99\columnwidth]{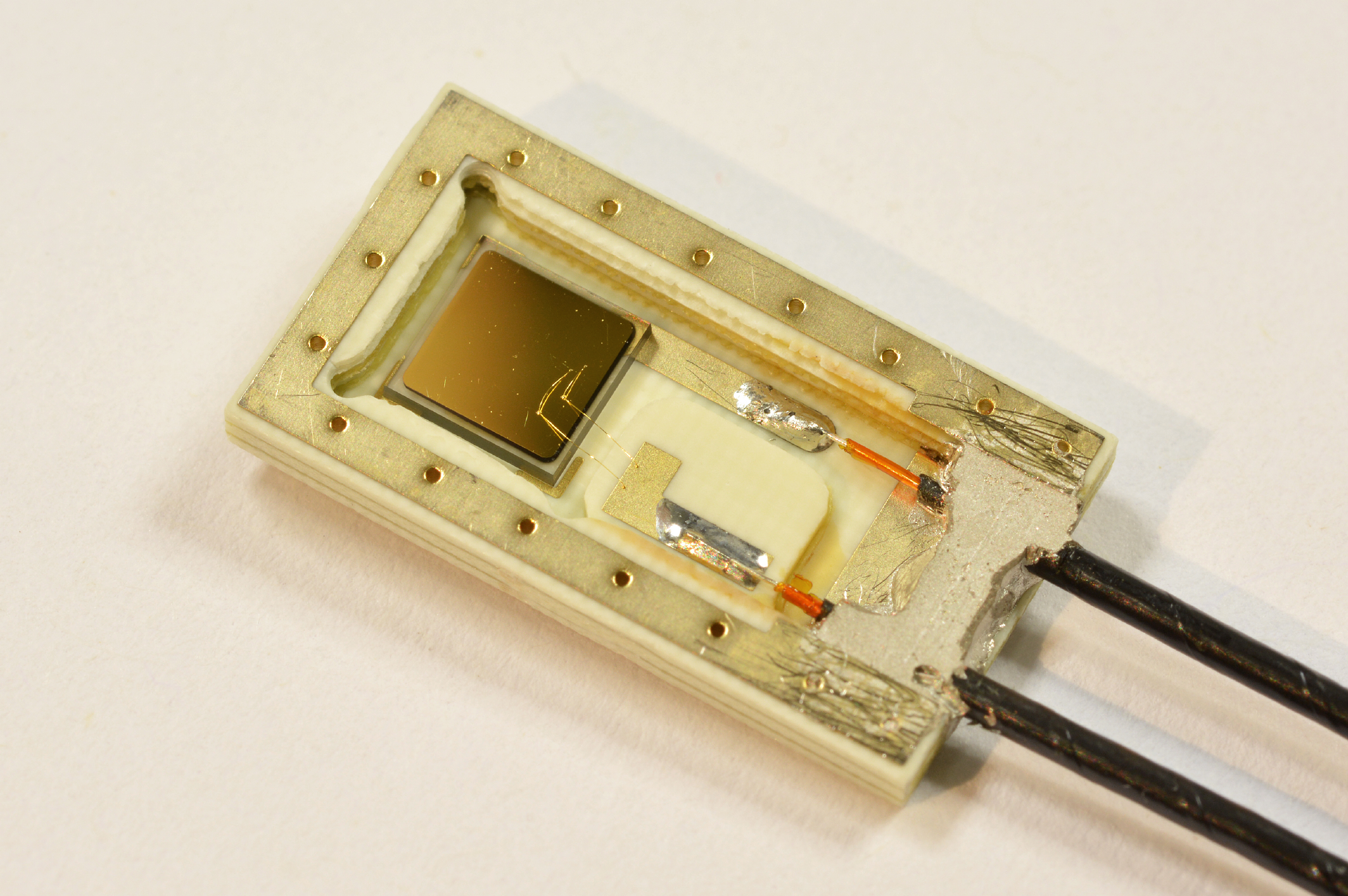}
		\put(86,57){\large (a)}
		\end{overpic}\\
		\vspace{2 mm}
		\begin{overpic}[width=0.99\columnwidth]{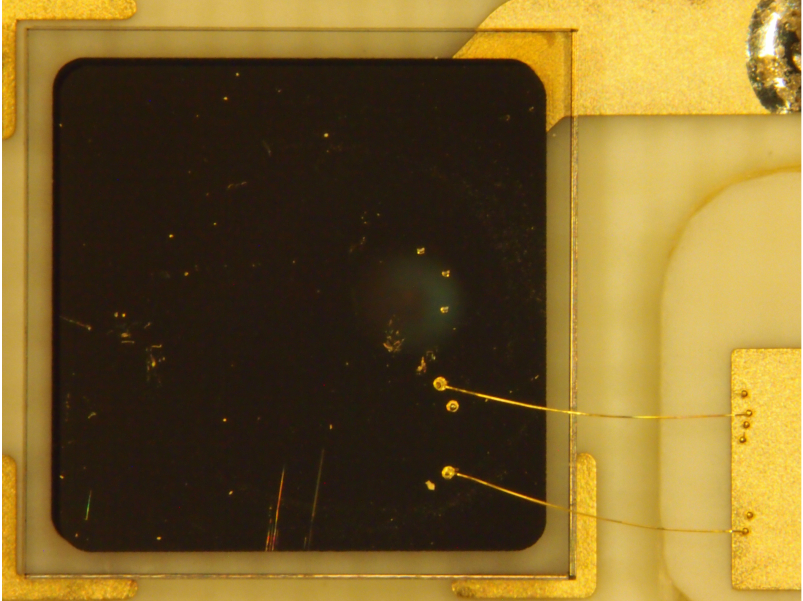}
		\put(86,65){\large (b)}
		\end{overpic}
	\caption{A diamond sensor packaged into a detector unit: (a) the sensor is glued on a Rogers printed-circuit board, with electrical contact established between the two electrodes (front/back) and the inner conductors of two miniature coaxial cables. (b) A zoom of the diamond sensor, with the front-side electrode connected to a soldering pad by two ball-bonded gold wires.}	
	\label{fig:diamonds_package}
\end{figure}

We mounted each diamond sensor in a ceramic-like Rogers printed-circuit board (PCB)~\cite{ref:rogers}, as shown in Figure~\ref{fig:diamonds_package}. 

We first soldered the inner conductors of two miniature coaxial cables, each $2.5$\,m long, to the printed board pads, while we used conductive glue to mechanically fix the cables and establish the connection of their outer conductors with the outer shielding of the printed-circuit board.  
This solution is mechanically more reliable (less prone to breaking of the cable shield following repeated bending) with respect to soldering the outer conductors of the coaxial cables to the package. The cable length was determined by constraints imposed by their use in the Belle~II setup.  

After cleaning the PCB-cables ensemble by isopropyl alcohol, using conductive glue we attached the diamond sensor to a square pad that connects the back-side electrode to the soldering pad of one of the coaxial cables. Finally, we connected the front-side electrode of the sensor to the soldering pad of the other cable by two ball-bonded gold wires.  
After the measurements presented in Sect.~\ref{subsec:TCT_alpha}, we completed the mechanical and electrical shielding  by gluing a thin ($180\,\upmu$m) aluminium cover on the front side of the package. 

We performed preliminary  measurements of dark current versus bias voltage  on all detectors as a quality control of the assembly process. We connected the diamond detectors to the measuring instrument via the miniature coaxial cables and, while keeping them in the dark, we scanned the bias voltage up to $\pm 800$\,V. 
Although we observed fairly large variations among different detectors, memory effects and long stabilization times, in all cases the measured dark current was less than about 10\,pA at $\pm 500$\,V and not larger than about $1$\,pA at $\pm 100$\,V. 

\section{Detector calibration}
\label{sec:detectors_calibration}
The response  of diamond sensors as dosimeters is not expected to be uniform. 
Crystal imperfections from the CVD-growing process can trap or recombine charge carriers generated by irradiation. Such imperfections differ from crystal to crystal, yielding  non-uniform charge-collection efficiency at a given bias voltage.   
Properties of the diamond-electrode interface are also sensor-dependent. Some detectors might feature non-blocking electrodes, which inject charge into the diamond bulk, and, in conditions that support the so-called photo-conductive gain, the charge-collection efficiency might even exceed unity~\cite{ref:diamonds_photoconductive_gain1,ref:diamonds_photoconductive_gain2}.

For these reasons, an individual calibration of each detector is needed in order to relate the measured current to a dose rate.  
Before the calibration, detailed in Sect.~\ref{subsec:beta}, we carried out two sets of measurements on each detector. 
The first set aimed at checking the transport properties of the charge carriers and 
the average ionisation energy to create an electron-hole pair.  
With this study, reported in Sect.~\ref{subsec:TCT_alpha}, we got an understanding of the homogeneity of the sensor properties within our sample. In the second set of measurements, we determined the suitable bias voltage for operating the sensors and checked the stability of the output current generated by irradiation. This study is presented in Sect.~\ref{subsec:detectors_stability}.

\subsection{Measurement of the charge-carrier properties}
\label{subsec:TCT_alpha}
To study the transport properties of electrons and holes, we used 
the transient-current technique (TCT)~\cite{ref:TCT}, which in our application employs monochromatic $\upalpha$~particles to generate electron-hole pairs localised at a small depth in the diamond bulk, very close to one  electrode. Depending on the bias polarity, charge carriers of one type are readily collected at the nearby electrode, while the others drift to the opposite electrode along the electric-field lines, inducing a current pulse. Information on the transport properties of the drifting carriers is determined by the features of this pulse: its shape is related to the carriers lifetime and to the uniformity of the electric field in the diamond bulk; its duration is related to the drift time  of the selected carriers; its integral to the collected charge. 

\begin{figure}[t]
	\centering
		\includegraphics[width=1.\columnwidth]{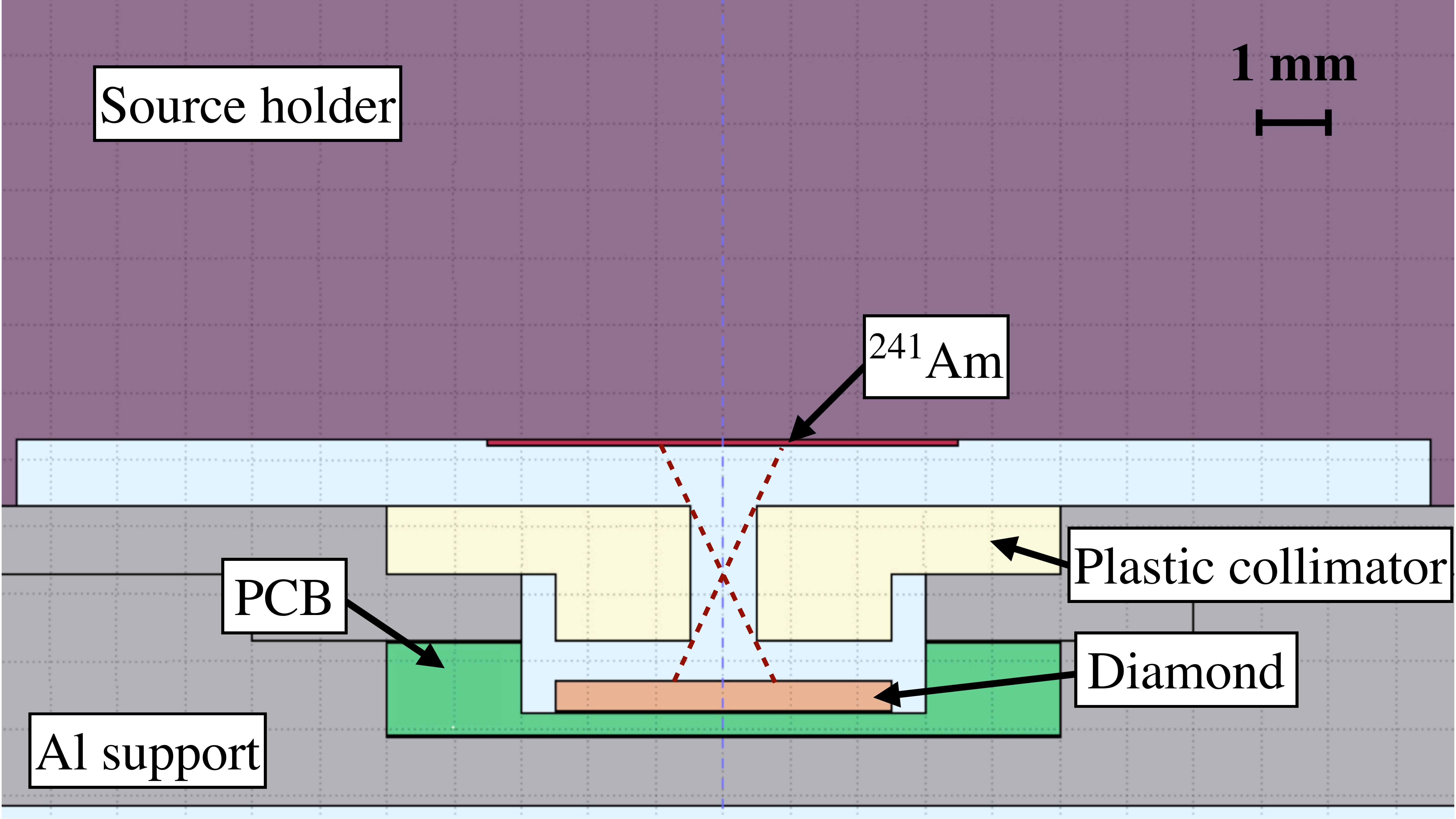}
	\caption{Schematic tranverse section  of the setup used to collimate the $\upalpha$ particles on the detector for  the TCT measurement. The $^{241}$Am is deposited on a circular surface of 7\,mm diameter. The red dashed lines show the maximum incident angles of the particle into the detector surface.}
	\label{fig:TCT_setup}
\end{figure}

To carry out the TCT measurements, we used a $5$-kBq $^{241}$Am source of $\upalpha$ particles, emitted with $5.485$\,MeV energy on average. The diamond detectors, assembled on their PCB package but without the aluminium cover, so that the $\upalpha$ particles could reach the detector surface, were inserted in an aluminium support. This support provided appropriate shielding and allowed to position the $\upalpha$ source at repeatable distance ($3.7$\,mm) from the front electrode of the diamond sensor, centered with respect to it. The $\upalpha$ particles were collimated by a Plexiglas insert, 2 mm thick, with a circular hole of 1 mm diameter, so that they would hit the sensor well within the electrode area, in a region where edge effects on the electric field can be neglected.
A scheme of the setup is shown in 
Figure~\ref{fig:TCT_setup}. 


From a detailed simulation of this setup based on the FLUKA software~\cite{ref:diamonds_FLUKA}, we estimated that an $\upalpha$ particle releases  in the diamond crystal $89.5\%$ of its energy,  corresponding to $4.91$\,MeV on average; only $3.9\%$ is released in the metallic electrode, while the rest is distributed between air and the collimating structure. The penetration depth in the diamond crystal is limited to about $12\,\upmu$m. 

We connected a high-voltage supply, delivering up to $\pm800$\,V, to  the back-side electrode of the diamond detector via a Bias-T circuit~\cite{ref:particulars} that, although not needed  to decouple the signal from the DC bias, has proven  effective in suppressing the noise. We connected the front electrode directly  to the input of a voltage amplifier with 3\,GHz band-width, 53\,dB gain and 50\,$\Omega$ input impedance (AM-02A from Particulars ~\cite{ref:particulars}). The output of the amplifier was analyzed by a digital storage oscilloscope. 
 
We optimised this experimental setup after a first set of measurements on a sub-sample of detectors, made with a less effective collimator, allowing the $\upalpha$ particles to hit the whole detector area (including the edges) with a wide range of incidence angles. In that first setup, we had also  used a different  readout scheme, yielding larger signal attenuation from cables and the Bias-T. Those differences have a negligible impact in the analysis of the charge-carrier transport, but significantly affect the estimated ionisation energy. We report the results obtained with the optimised setup and note the distinction from the old configuration only when it is relevant. 

\begin{figure}[t]
	\centering
		\includegraphics[width=0.9\columnwidth]{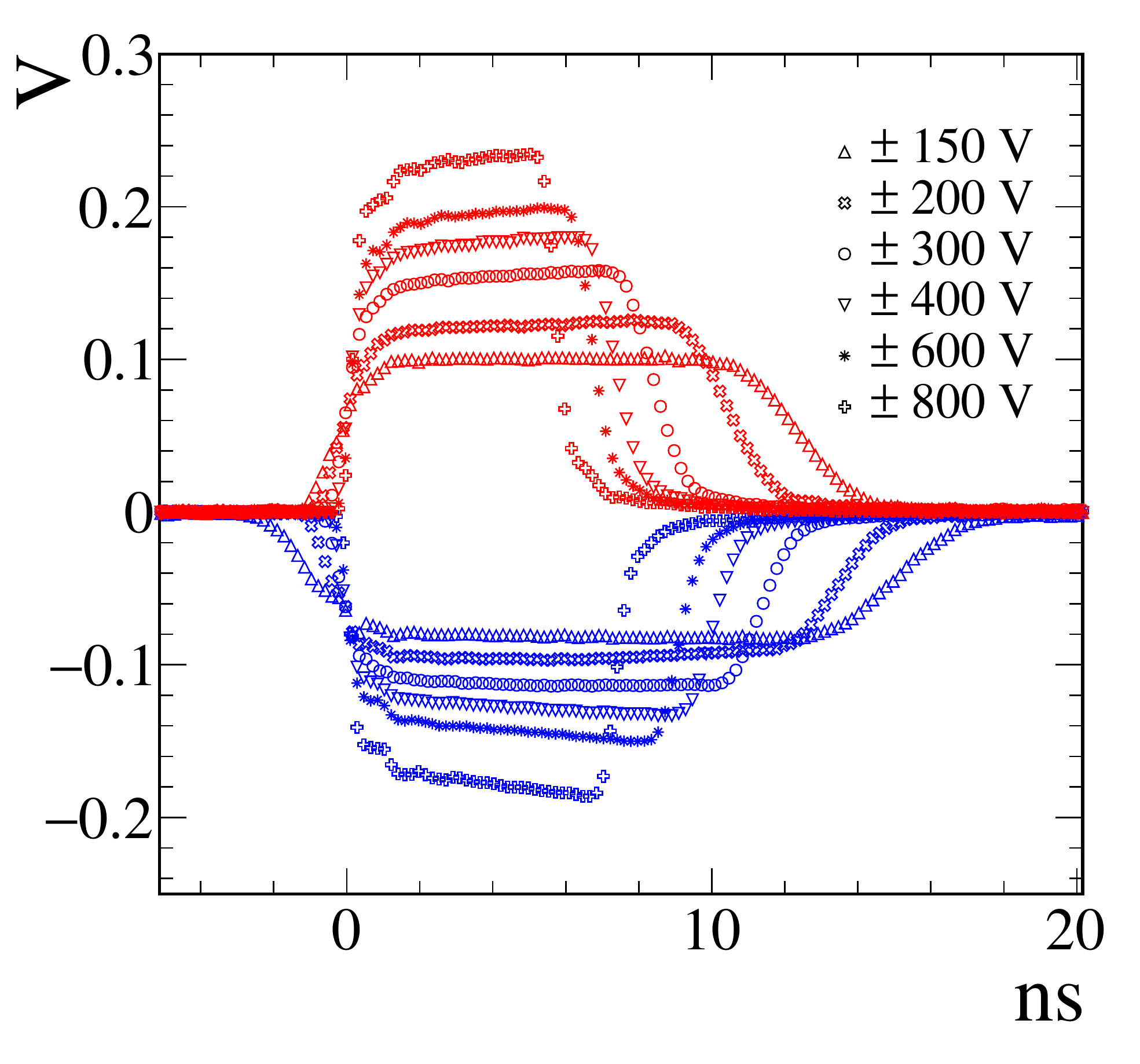}
	\caption{A summary of the time development of the signals (averages of 1000 signal pulses) induced by electrons (negative signals, blue markers) and holes (positive signals, red markers), for the different bias voltages listed in the legend. The sign of the signal induced by electrons (holes) is inverted by the amplifier.}	
	\label{fig:TCT_signal}
\end{figure}

We analysed the signal generated by the  $\upalpha$ particles for twelve choices of the bias voltage between $\pm 800$\,V. 
We discarded measurements with bias voltage between $\pm 150$\,V, 
because the trigger level of the oscilloscope, kept above noise, biased the measurement for lower pulses.
Figure~\ref{fig:TCT_signal} reports the time development of the average of 1000 signal pulses  from a diamond detector.  
Each measurement provides the average pulse-integral $A$ and the average pulse-width at half height $w$. We assumed a uniform electric field in the diamond bulk (as suggested by the approximately flat top of the signal shapes) and a drift distance of the carriers equal to the sensor thickness $h = 0.50$\,mm. For each value of the electric-field intensity $\mathcal{E}$, we estimated the drift velocity $v_{\rm drift} = h/w$ and the mobility $\upmu = v_{\rm drift}/\mathcal{E}$ for both drifting carriers.  

\begin{figure}[t]
	\centering
	\includegraphics[width=0.9\columnwidth]{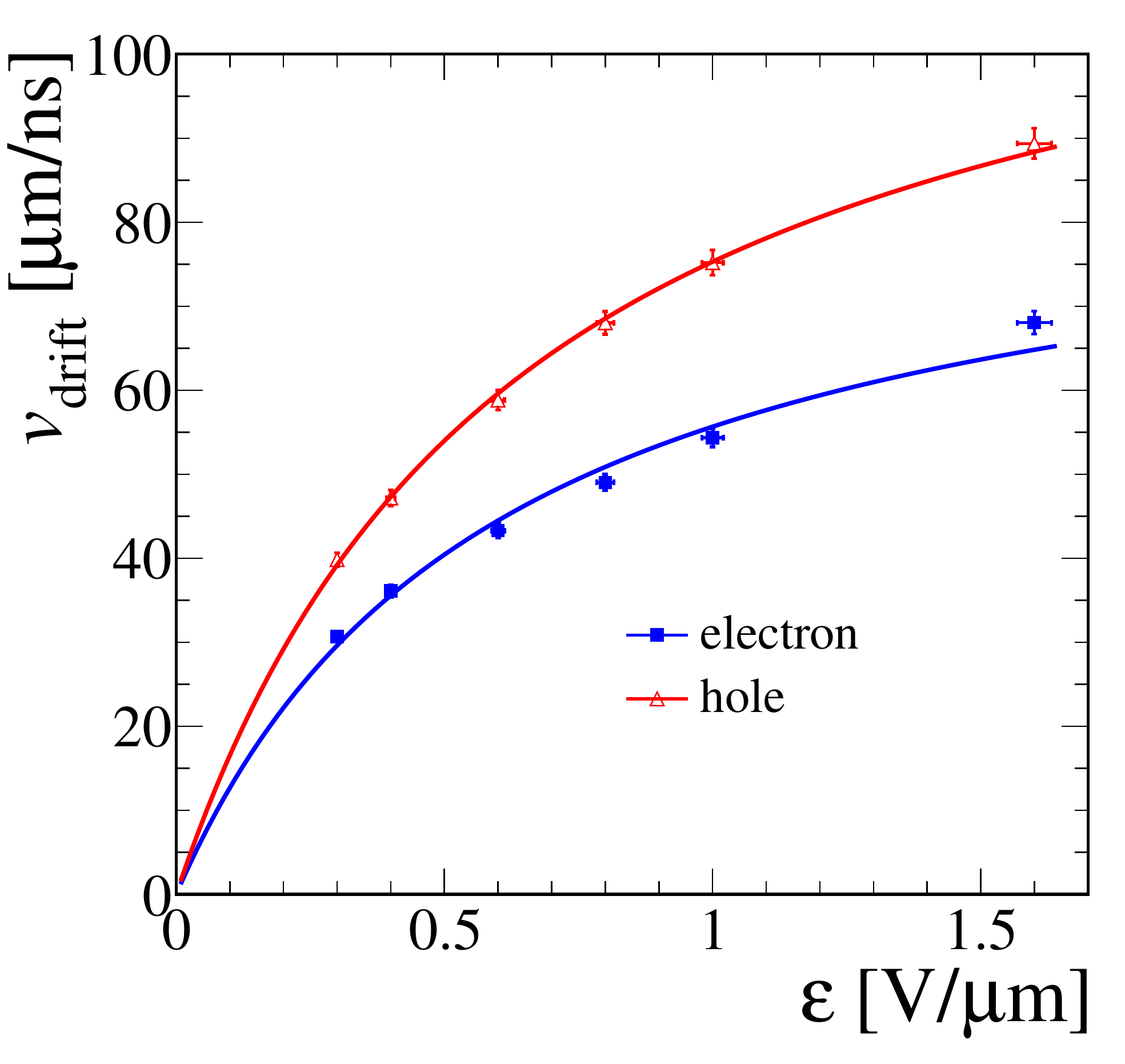}
	\caption{Example of the measured drift velocity $v_{\rm drift}$ as a function of the electric-field intensity $\mathcal{E}$, for (solid markers) electrons and (open markers) holes. Data points include the uncertainties on $v_{\rm drift}$ and the electric field intensity $\mathcal{E}$ from propagating the uncertainty on the sensor thickness, the uncertainty on the average pulse-width of each signal, and that on the voltage bias. 
	 The lines represent a fit with Eq.~\eqref{eq:velocity}.}	
	\label{fig:velocity}
\end{figure}

We empirically described the drift velocity as a function of  $\mathcal{E}$ by the following expression~\cite{ref:TCT}:
\begin{equation}
\label{eq:velocity}
v_{\rm drift}(\mathcal{E}) = \frac{\upmu_{0}\,\mathcal{E}}{1+ \frac{\upmu_{0}\,\mathcal{E}}{v_{\rm sat}}}\,,
\end{equation}
where $\upmu_{0}$ is the mobility extrapolated to low field-intensity, and $v_{\rm sat}$ is the saturation velocity at high field. Figure~\ref{fig:velocity} shows an example of $v_{\rm drift}$ as a function of $\mathcal{E}$ with a fit of the data using Eq.~\eqref{eq:velocity}. 
The values of $\upmu_{0}$ and $v_{\rm sat}$ obtained by fitting Eq.~\eqref{eq:velocity}, averaged over our sample of diamond detectors, are: $\overline{\upmu}_{0} = 2.0 \times 10^{3}\,\mathrm{cm^2/Vs}$ and $\overline{v}_{\rm sat}=1.3 \times 10^{7}\,\mathrm{cm/s}$ for holes, $\overline{\upmu}_{0} = 1.7 \times 10^{3}\,\mathrm{cm^2/Vs}$  and $\overline{v}_{\rm sat}=0.9 \times 10^{7} \,\mathrm{cm/s}$ for electrons. The standard deviation of these results is about $20\%$, confirming that the transport properties of charge carriers are sufficiently homogeneous. We did not observe significant differences between the results obtained with the optimised and non-optimised setup. 

 For each pulse integral $A$ measured by the oscilloscope, we estimated the collected charge $Q$   as
$Q = A/ (Z_{\rm in}\,G_{\rm amp})$, 
where $Z_{\rm in}$ and $G_{\rm amp}$ are the input impedance and the measured gain of the amplifier, respectively. 
Independently of the setup used, we observe that the estimated collected charge versus  applied voltage saturates 
to an approximately constant value. 
The fractional difference between the collected charge evaluated from drifting electrons or  holes does not exceed $2\%$.   

From the collected charge $Q$, we determined the average ionisation energy $E_{eh}$ for electron-hole pairs as
\begin{equation}
E_{eh} = \frac{q_{e}\,E}{Q}\,,
\end{equation}
where $E$ is the average energy released in diamond by the $\upalpha$ particles and $q_{e}$ is the charge of the electron. 
We computed the energy $E$  from a detailed simulation of the source, the collimator, and the sensor with its support. We did not simulate the charge-carrier transport and we assumed a full charge-collection efficiency. 
We obtain an average value $E_{eh} = 13.1$\,eV and $E_{eh} = 16.3$\,eV with the optimised and non-optimised setup, respectively.   
The results obtained with the non-optimised setup are understood to originate from a biased estimation of the collected charge, due to attenuation in circuit components, to long tails in the shape of the  signals, and a contamination of about 10\% of signals with irregular shape, caused by a poor collimation of the $\upalpha$ particles. Taking into account those biases, all values are consistent within 5\% with an average value of $13$\,eV, which is usually quoted in  literature for high-quality diamond sensors~\cite{ref:Eleni}.

\subsection{Current-voltage profiles under irradiation and current stability}
\label{subsec:detectors_stability}
A stable and reproducible response to radiation is crucial for dosimetry and monitoring. The detector response might depend on the applied bias voltage, and the optimal operation value has to be determined. We obtained this  by measuring the current-voltage ($I$-$V$) profile under irradiation for each detector after the full assembly ({\it i.e.} after gluing  the aluminium cover on the PCB package). 

We measured the current over a $\pm 500$\,V bias-voltage range. We irradiated the detectors  for 2--5 minutes for each value of the bias voltage. The radiation was provided by a steady flux of $\upbeta$ electrons from the source described in Sect.~\ref{subsec:beta}, using the same setup detailed therein. We placed the source at a fixed distance, which ranged from about $2$\,mm to $4$\,mm from the detector, yielding currents typically between 1--2\,nA at  $\pm 100$\,V~bias. 

\begin{figure}[t]
	\centering
    	\begin{overpic}[width=0.49\columnwidth]{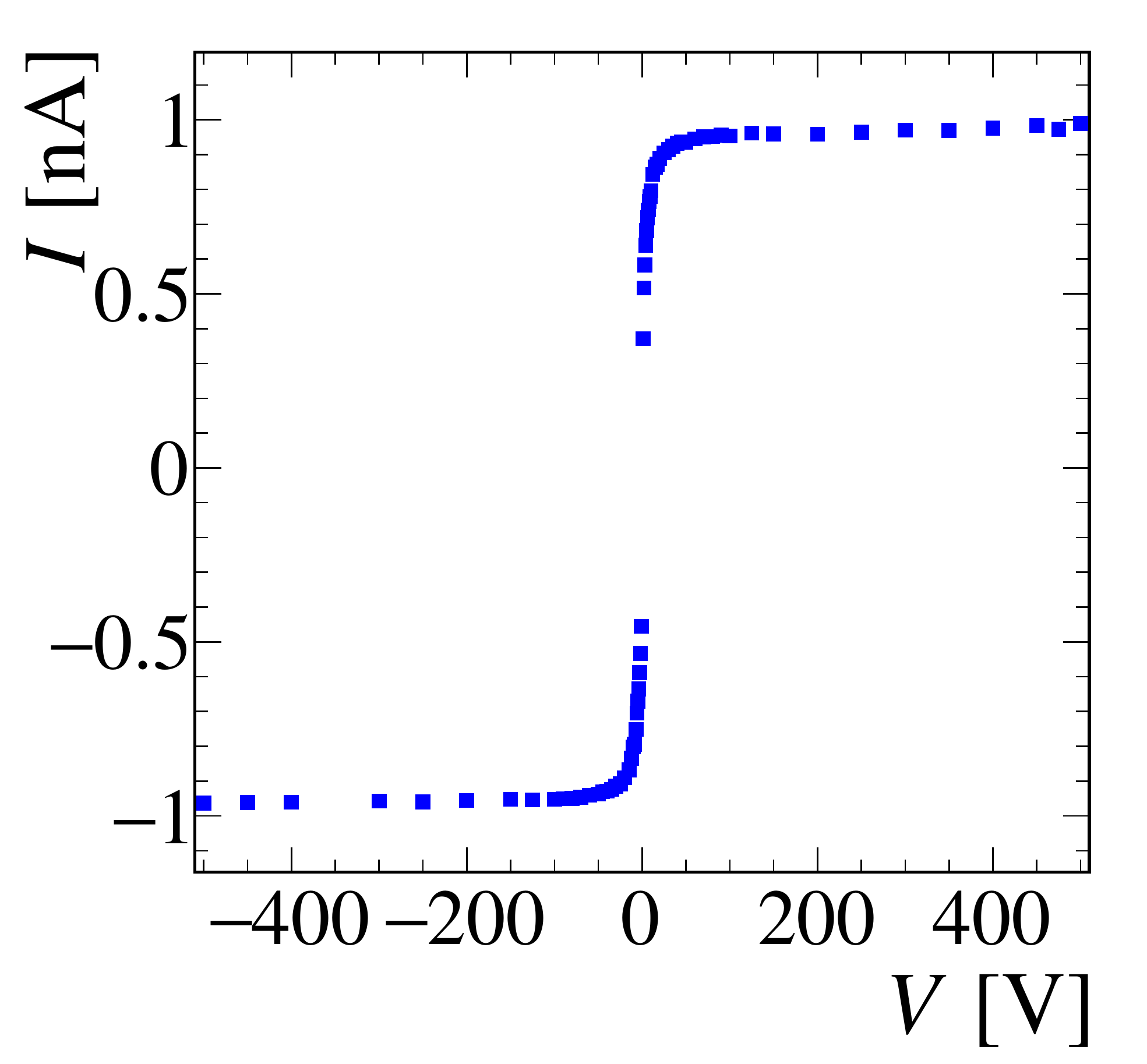}
    	\put(78,25){(a)}
    	\end{overpic}
    	\begin{overpic}[width=0.49\columnwidth]{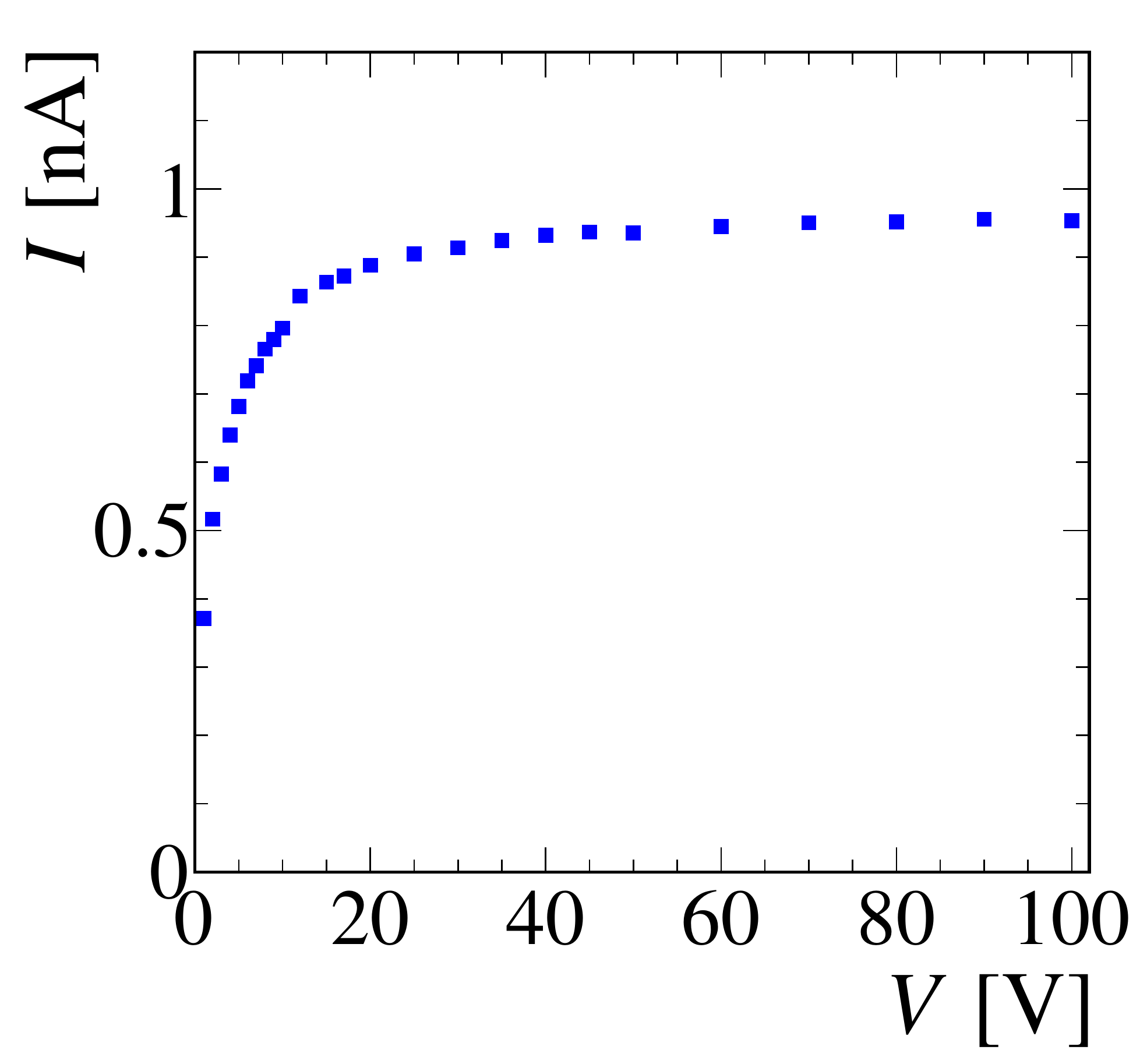}
    	\put(78,25){(b)}
    	\end{overpic}\\
		\begin{overpic}[width=0.49\columnwidth]{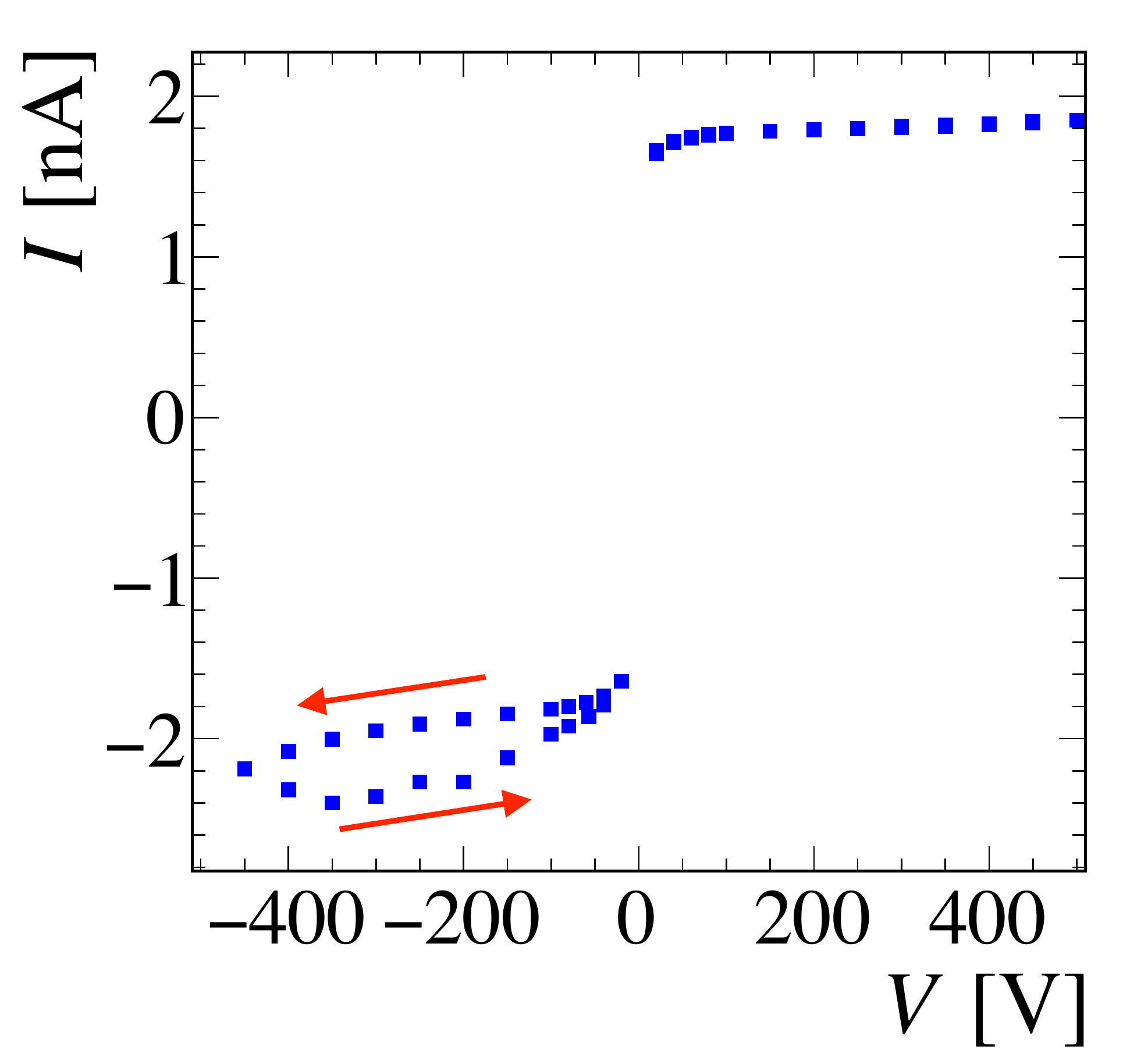}
		\put(78,25){(c)}
		\end{overpic}
		\begin{overpic}[width=0.49\columnwidth]{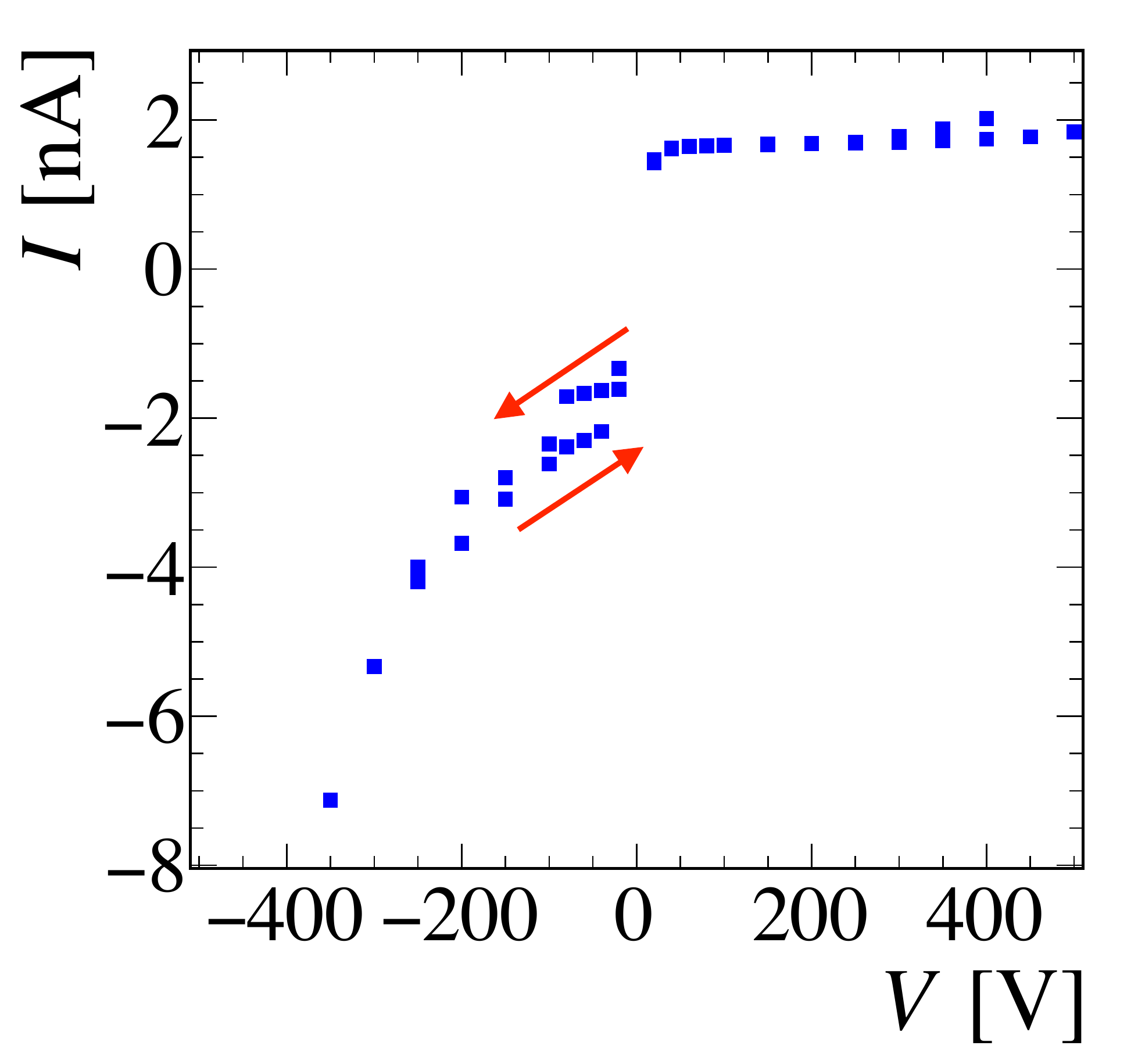}
		\put(78,25){(d)}
		\end{overpic}
	\caption{Examples of $I$-$V$ measurements from three diamond sensors.  (a) The $I$-$V$ profile is symmetric around $0$\,V for bias voltages of opposite polarity, and the current reaches a stable plateau for $|V|\approx 60$\,V; this is a typical situation for half of our sensors. (b) A zoom in the range  $0$--$100$\,V shows the sharp rise of the current in a couple of volt. (c-d) Two sensors with an asymmetric $I$-$V$ profile, exhibiting also hysteresis effects for negative voltages (the red arrows indicate the time order of the measurements). (d) A small hysteresis effect is also present for the best polarisation, but variations of the current are much smaller than those on the opposite side, and are significant only at high voltages.}	
	\label{fig:IV_profile}
\end{figure}

Figure~\ref{fig:IV_profile} shows some examples of the $I$-$V$ profiles obtained from three sensors exhibiting different behaviour. Plot (a) represents the situation typical for about a  half of our sensors, which display a symmetric $I$-$V$ characteristic. In these cases, the current reaches a plateau for $|V|$ between $60$--$80$\,V, remaining approximately flat up to $|V| =  500$\,V. Plots~\ref{fig:IV_profile} (c) and (d) are examples of $I$-$V$ profiles characterising the remaining sensors, where an asymmetric response is observed. For one polarity the current reaches a constant value for $|V|$ typically greater than $80$\,V, while for the opposite polarity no current saturation is observed. 

For those detectors with an asymmetric $I$-$V$ profile, the response can also  present an 
hysteresis: the current value measured at a given voltage depends on the direction of the voltage scan, indicating that the current is not a function of $V$ alone, but depends also on the previous ``hystory''. The variation of the current due to such hysteresis effects is not significant at $|V|=100$\,V for the bias polarity  showing the plateau, although  it  can be much larger in the opposite polarity. 

We cannot provide a clear explanation for the asymmetric $I$-$V$ profiles and the hysteresis effects; possible hypotheses range from a crystal asymmetry from the growth process to features of the diamond-electrode interface. For these sensors we decided to adopt the voltage polarity leading to a saturation of the current.

For a couple of sensors, we also sampled the region $0$--$100$\,V in finer steps, to study in more detail the onset of charge-collection efficiency. An example is shown in Figure~\ref{fig:IV_profile} (b). With a bias voltage as small as $2$\,V, the measured current is about 50\% of the plateau value. The charge-collection efficiency rises very steeply within a couple of volt, and  slowly saturates increasing the bias. 

We chose $\pm100$\,V as the optimal operation voltage  (with the sign set for each sensor, according to the $I$-$V$ profiles). With this choice, our detectors have a (steady-state) charge-collection efficiency close to 100\% on average. 

\begin{figure}[t]
	\centering
    	\begin{overpic}[width=0.49\columnwidth]{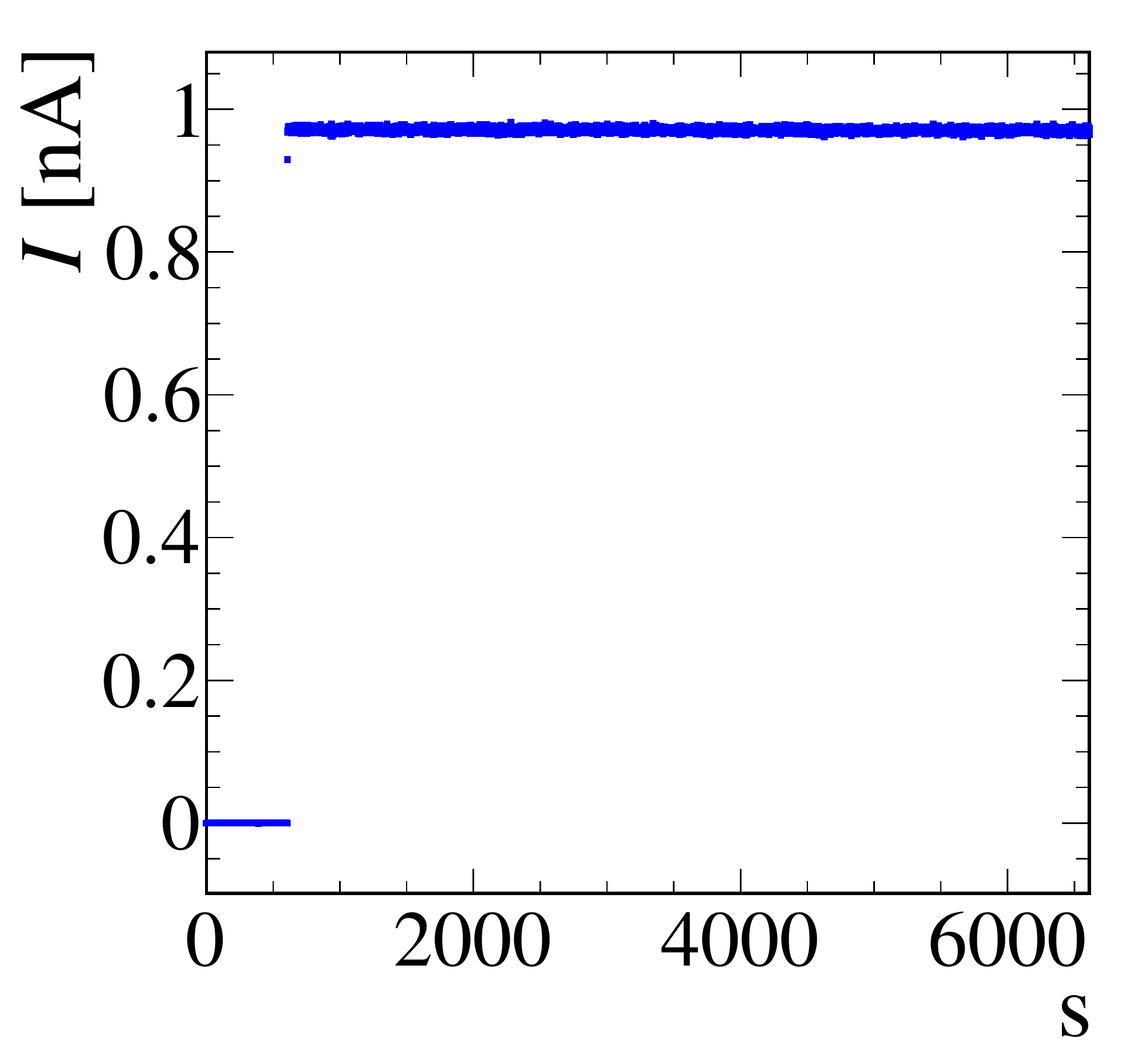}
    	\put(80,21){(a)}
		\end{overpic}
    	\begin{overpic}[width=0.49\columnwidth]{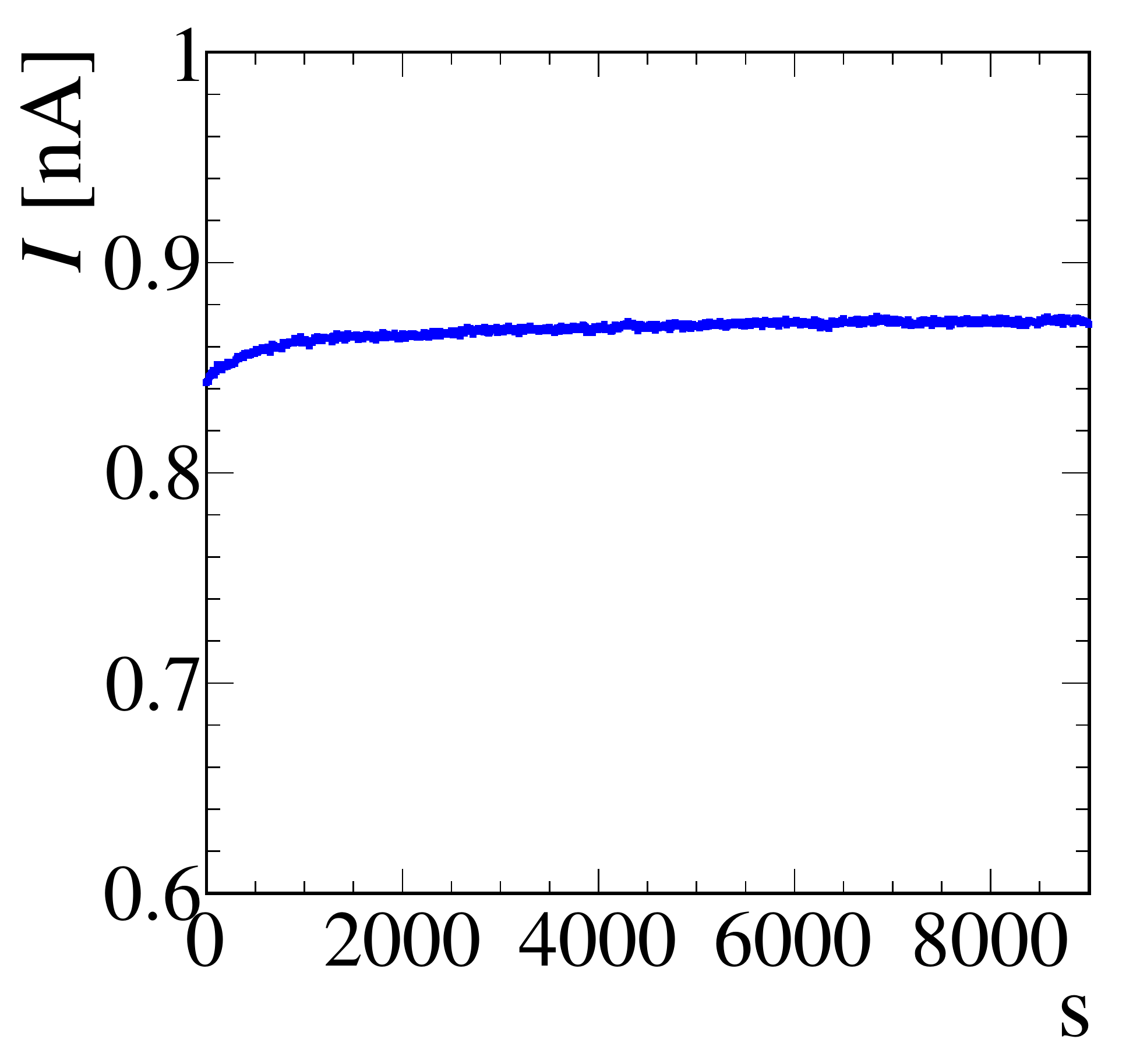}
    	\put(80,21){(b)}
		\end{overpic}\\
    	\begin{overpic}[width=0.49\columnwidth]{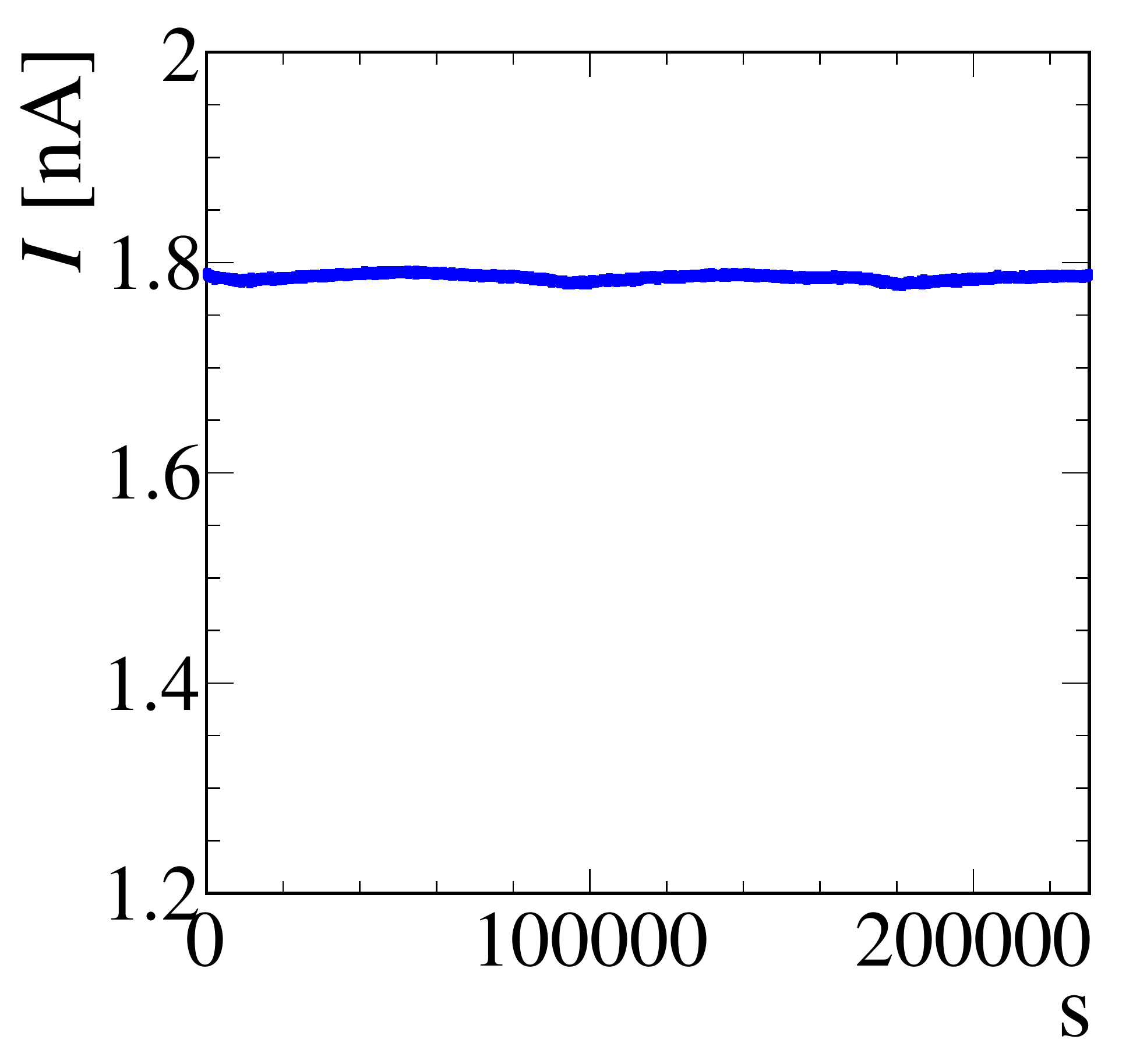}\put(80,21){(c)}
		\end{overpic}
       	\begin{overpic}[width=0.49\columnwidth]{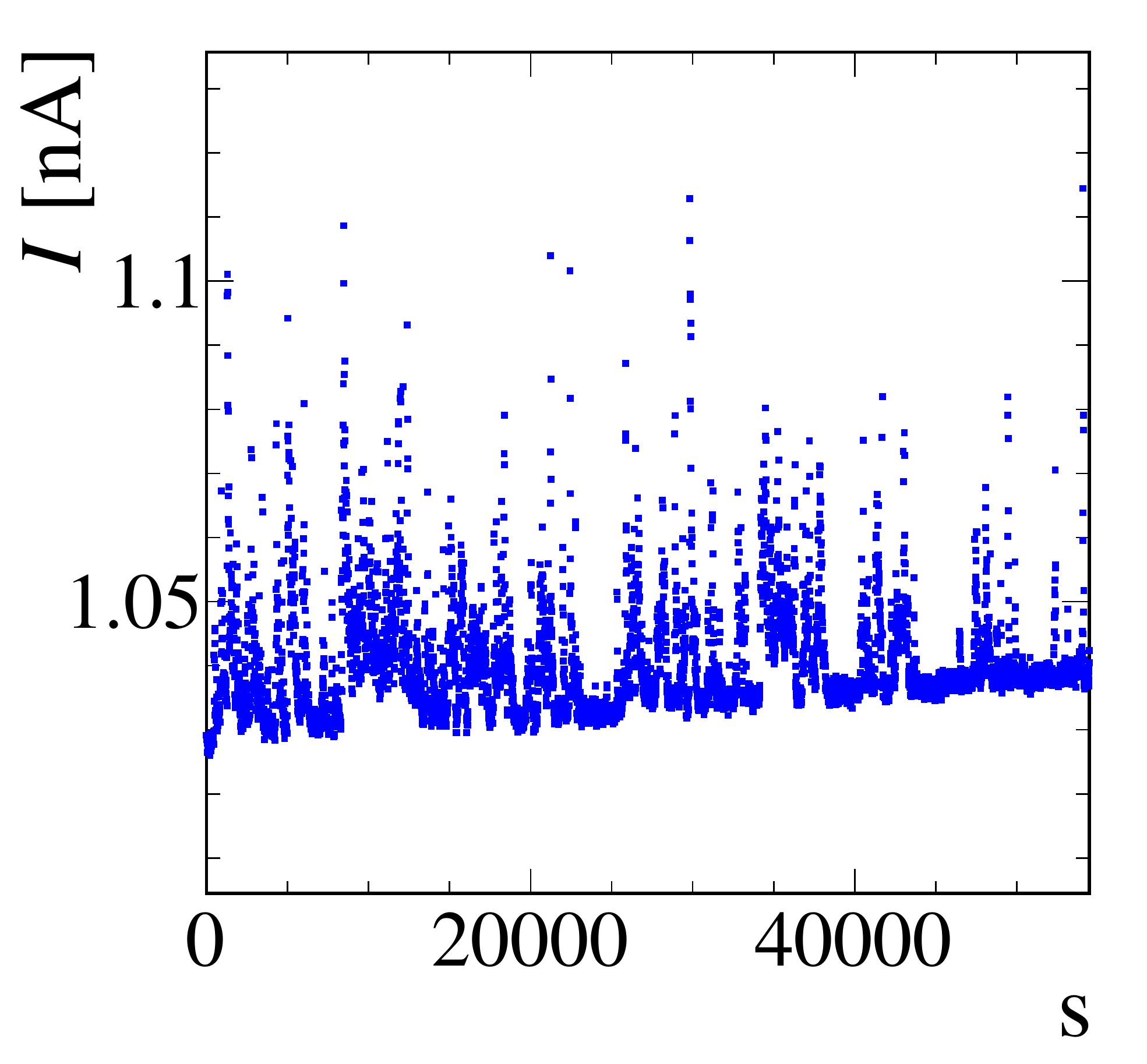}
       	\put(80,21){(d)}
		\end{overpic}
	\caption{Examples of current-stability measurements for a couple of sensors, with a $|V|=100$\,V bias applied (choosing the best polarity for the sensors with asymmetric $I$-$V$ profiles). (a) In the first $25$\,minutes, a shutter is placed in front of the detector to stop completely the radiation; the dark current is constant at a very low value ($<0.1$\,pA in this case). When the shutter is removed, the current quickly reaches ($<1$\,s) a stable value and then remains constant. (b) For some detectors (about $10\%$ of our sample), the current reaches about 90\% of its stable value within 1\,s, with a longer transient time to the asymptotic value. (c) Stable output currents are generally observed also over a couple of days; a small periodic variation of the current in the plot is correlated with  day-night temperature changes. (d) For detectors featuring an asymmetric $I$-$V$ profile, on the worst bias polarity the current can show large and sudden variations: in this example we observe random spikes with variations of the order of $10\%$ for a current of about $1$\,nA.}	
	\label{fig:instab_transient}
\end{figure}

We checked the stability in time of the radiation-induced current at the chosen operation voltage, over a time span ranging from several hours to a few days.  Figure~\ref{fig:instab_transient} reports a few typical examples. On the sensors exhibiting symmetric $I$-$V$ profiles, we observed a very stable response over time with values constant within $1\%$. For those with asymmetric $I$-$V$ profiles, we observed constant current values  for the best polarity, with  fluctuations no larger than 5\%. We noticed larger instabilities, also with random spikes, for the opposite polarity. These stability tests confirmed our choice of the best polarity for the sensor with asymmetric $I$-$V$ profile.  

We also observed a different time response between the sensors. Although all detectors react within 1\,s (time-resolution of the current sampling in these measurements) from the start of the irradiation, some of them present a two-step  response: about 90\% of the value of the current is reached within 1\,s,  followed by a slower transient to 100\% with time duration ranging from tens of seconds to many minutes. We attribute this behaviour to different amounts of crystal defects, corresponding to trapping and de-trapping effects of the charge carriers following irradiation transients~\cite{ref_Bergonzo}.  

\subsection{Current-to-dose calibration factors}
\label{subsec:beta}
To employ the diamond detector as a dosimeter, we need to relate the current measured under irradiation to a dose rate. We define a calibration factor $k$, such that
\begin{equation}
\frac{dD}{dt}  \equiv  \frac{1}{m} \frac{dE}{dt} = k \, I\,,
\end{equation}
where $dD/dt$ is the dose rate, {\it i.e.} the energy per unit  time $dE/dt$ released by the radiation in the detector mass $m$, and $I$ is the measured current.

To account for non-uniform response of our detectors, we split $k$ into two terms, as 
\begin{equation}
\label{eq:k100_def}
k \equiv \frac{F}{G}\,, 
\end{equation}
where $F$ is a constant factor that takes the value
\begin{equation}
F\equiv \frac{E_{eh}}{m\, q_e} = 34.9\,\mathrm{(mrad/s)/nA}\,,
\end{equation}
considering the mass of our detectors $m = 37$\,mg and an average ionisation energy to create electron-hole pairs $E_{eh}= 13$\,eV. 
The factor $G$ is a dimensionless constant characteristic of each detector. A value $G=1$ would represent (i) a fully efficient detector, 
(ii) with blocking electrodes, and (iii) ionisation energy of 13\,eV. Any deviation from unity is a measurement of the departure from  the assumptions (i)--(iii). We notice that assumption (ii) implies no amplification factor from the ``photoconductive gain" process~\cite{ref:diamonds_photoconductive_gain1,ref:diamonds_photoconductive_gain2}.   

To determine the value of $G$, we used a silicon diode as a reference. We exposed the diamond detectors and the diode to a source of $\upbeta$ radiation, and we determined the ratio of the \emph{signal} current from the diamond detector to the \emph{reference} current from the diode. We obtain the characteristic constant $G$  by comparing the measured \emph{signal/reference} current ratio and that expected from a detailed simulation of the experimental setup, where we adopt the hypotheses (i)--(iii) for the diamond detectors  and we assume a  good knowledge of the diode response. The use of a reference greatly reduces uncertainties associated to the source activity and to the simulation of the setup, that would otherwise limit the accuracy of the calibration procedure.

\subsubsection{Measurement of signal-to-reference ratio of currents}
\label{subsubsec:beta_setup}
We exposed the diamond detectors  to a steady flux of $\upbeta$ electrons emitted by a $^{90}$Sr radioactive source. We moved the source  along a straight line orthogonal to the detector surface and traversing its center, and changed the source-detector using  a stepper motor driven by an Arduino micro-controller~\cite{ref:arduino}. 
The origin is known with an accuracy of $0.2$\,mm. 

The $\upbeta$ source  consists of an ion-exchange organic spherical bead, $1$\,mm in diameter, with radioactive nuclei uniformly distributed in the volume. The bead is mounted on top of a steel needle embedded in a Plexiglas container. Electrons are emitted isotropically from the $\upbeta$ decays $\mathrm{^{90}Sr} \to \mathrm{^{90}Y}\, e^- \overline{\nu}_e$ and $\mathrm{^{90}Y} \to \mathrm{^{90}Zr}\, e^- \overline{\nu}_e$, with a known energy spectrum up to about $2$\,MeV.  
Being the $^{90}$Y half-life 
much shorter than that of $^{90}$Sr, 
the $^{90}$Y decay rate is {\it in equilibrium} with its production rate. Thus, the rate of $\upbeta$ electrons emitted from the source is twice the $^{90}$Sr activity. At the time of these measurements, the activity of the source from the $^{90}$Sr decays was approximately 3\,MBq. 

\begin{figure}[t]
	\centering
		\begin{overpic}[width=0.9\columnwidth]{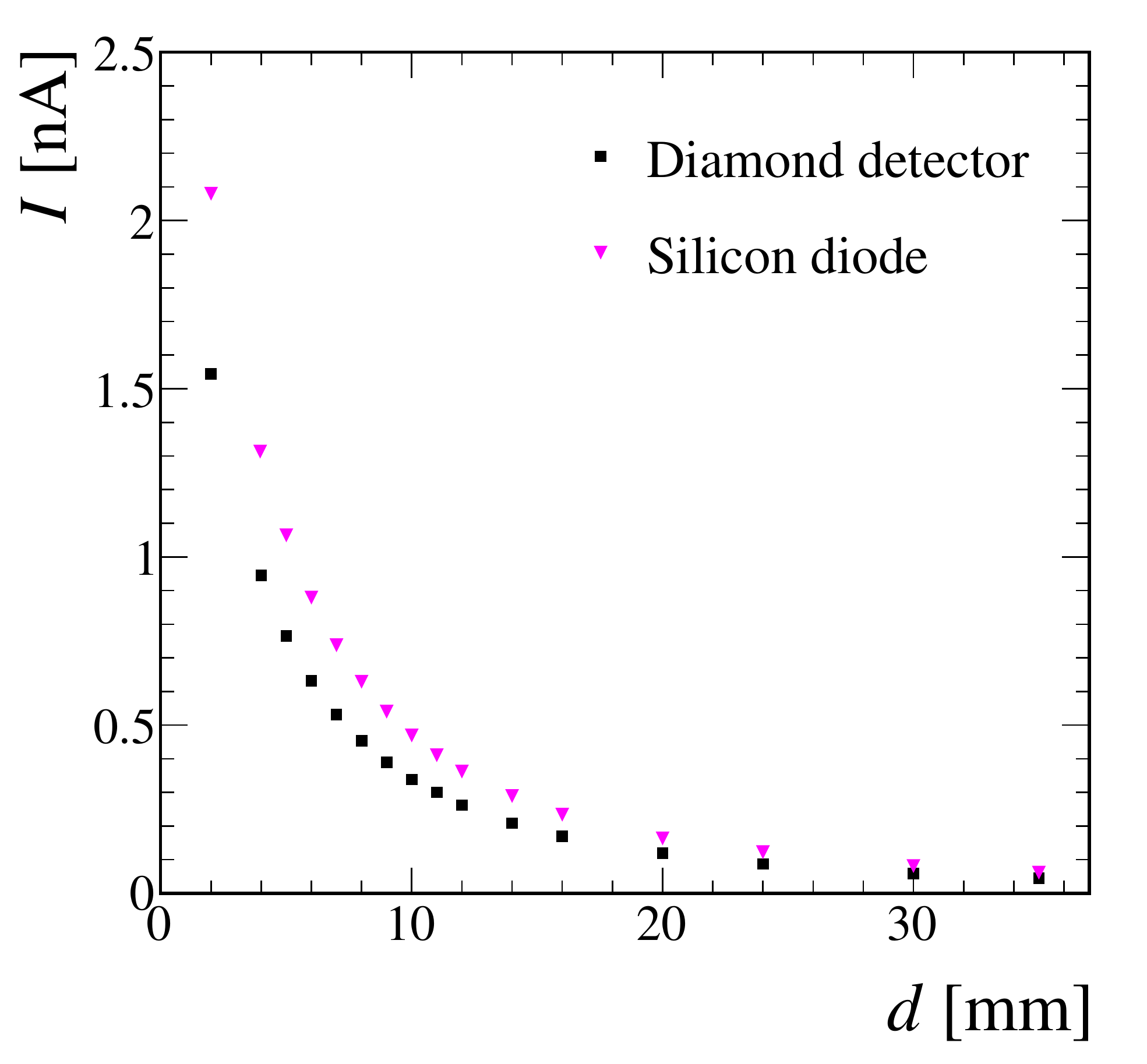}
		\put(80,30){\large (a)}
		\end{overpic}\\
		\begin{overpic}[width=0.9\columnwidth]{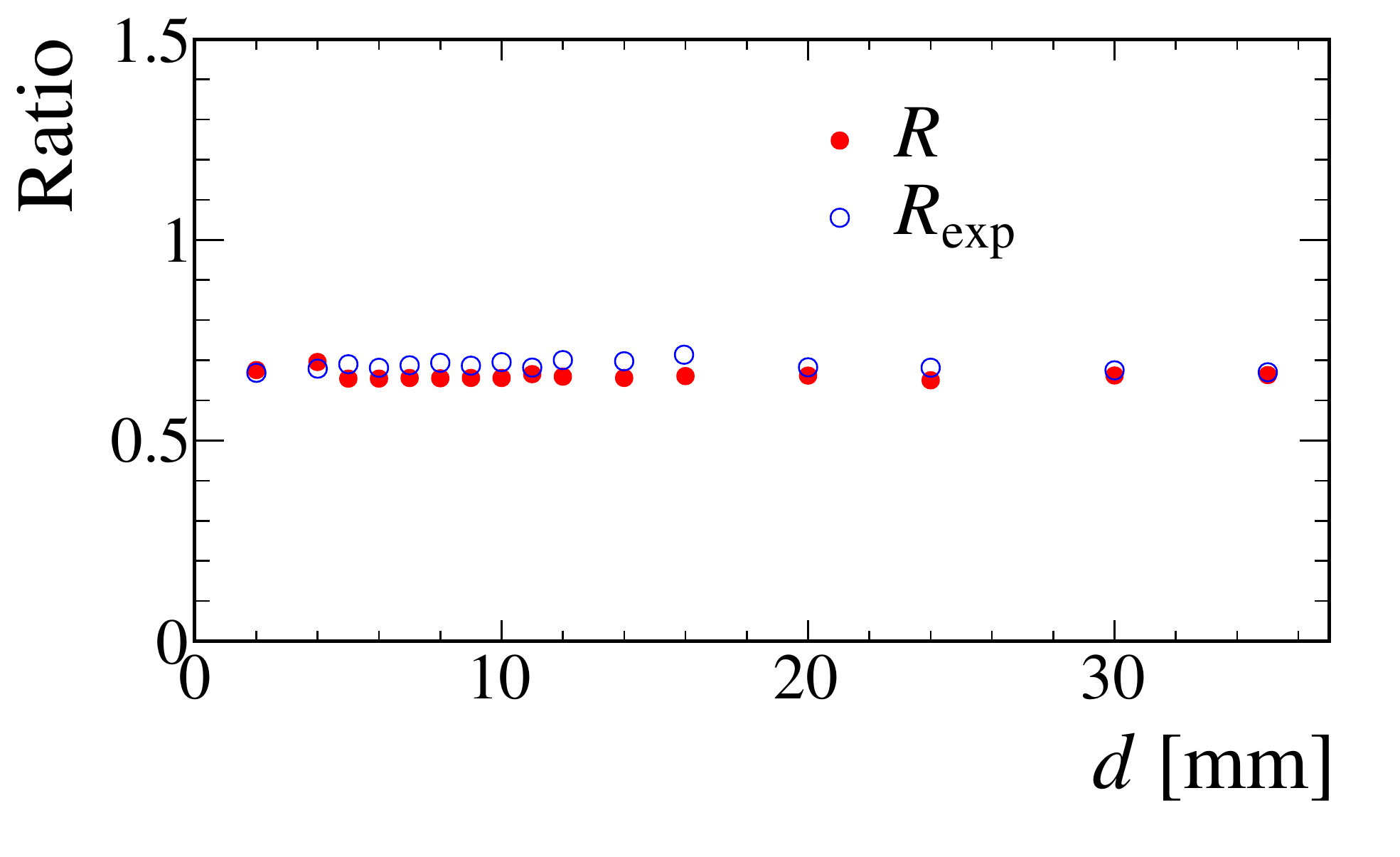}
		\put(80,22){\large (b)}
		\end{overpic}
	\caption{Current versus source-detector distance. (a)  Example of (square markers) the signal current $I$ measured from a diamond detector, and (triangular markers) reference current $I_{\rm r}$ measured from the silicon diode. (b) The signal-to-reference ratio of currents $R$ compared to the expected ratio $R_{\rm exp}$ from Eq.~\eqref{eq:R_exp}. In both plots, the  uncertainties are not visible because of their small size compared to that of the markers.}	
	\label{fig:I_vs_d}
\end{figure}

We connected the detector  to a bias-voltage supply~\cite{ref:CAEN_HV} and to a pico-ammeter~\cite{ref:Elettra_AH501} which measured the  signal current.   
We interfaced the stepper motor, the voltage source, and the pico-ammeter to a computer to automatically perform complete sequences of measurements.
We explored a range of source-detector distance $d$ from $2$\,mm to $35$\,mm and for each distance we recorded the signal current $I(d)$, averaging $2.8$ million readings taken in 2 minutes. An example is shown in Figure~\ref{fig:I_vs_d}.

We used the same setup and procedure to measure the reference current $I_{\rm r}(d)$ from the silicon diode. The diode features an n bulk, a p$^{+}$ layer obtained by Boron ion implantation, and a n$^{+}$ layer for the ohmic contact. The square-shaped diode, fabricated on a substrate $25.0\pm0.1$\,mm$^2$ wide and $0.455\pm 0.010$\,mm thick, is hosted in a package similar to that of the diamond detectors.  The diode is closely surrounded by a p$^{+}$  guard ring which delimits the charge collection volume of the central diode, given by an effective area of $12.2 \pm 0.2$\,mm$^2$ and the depletion thickness. We applied a  100\,V bias  to the n$^{+}$ contact, with the p$^{+}$ layer and connected the guard ring  to ground. In this condition, the diode was overdepleted with an active volume given by the diode effective area and the substrate thickness. 

 While for diamond detectors the dark current was negligible, it contributed significantly in the measurements with the diode, especially at far source-detector distance. Thus, each time we changed the distance, we monitored the dark current, and we subtracted its value from the current measured under irradiation. The dark current was constant at about $0.01$\,nA. The reference current from the diode  measured as a function of $d$ is shown in Figure~\ref{fig:I_vs_d}.

Both signal and reference currents are roughly proportional to $d^{-2}$, as expected from the variation of the $\upbeta$-electron flux on the detector surface as a function of $d$, for an almost-pointlike source.  
The signal-to-reference ratio of currents, 
\begin{equation}
\label{eq:R_meas}
R \equiv \frac{I(d)}{I_{\rm r}(d)}\,,    
\end{equation}
is constant versus $d$, as shown in the Figure~\ref{fig:I_vs_d} (b). The flatness of $R$ denotes that the measured values of the  source-detector distance are compatible in the two sets of data.  

\subsubsection{Computation of the expected ratio of currents}
\label{subsubsec:montecarlo}
We computed the expected signal (reference) current $I_{\rm exp}$ ($I_{\rm exp}^{\rm \, r}$) from the energy released per $\upbeta$ electron in the detector volume, $E$ ($E_{\rm r}$).
We used the FLUKA software~\cite{ref:diamonds_FLUKA} to obtain the released energy, by implementing a realistic and detailed model of the full setup, from the spherical source to the detectors, with all supports and packaging, to account for the full geometry and all materials traversed by the radiation.      

We did not attempt to model the charge-carrier generation and transport. 
 We used a simplified model where all energy released in the active volume generates charge carriers. For the diamond detector, we used assumptions (i)--(iii) presented in Sect.~\ref{subsec:beta}; for the diode, we assumed an average ionisation energy $E_{eh}^{\,\rm r} = 3.66\pm 0.03$\,eV~\cite{ref:SiEeh} and a full charge-carrier collection efficiency. In this model, the expected current is expressed as
\begin{equation}
\label{eq:estimated_current}
I_{{\rm exp}}^{\,\rm (r)}(d) = 2 \,\mathcal{A}\, \frac{E_{\rm (r)}(d)}{E_{eh}^{\, \rm (r)}}\,q_{e}\,,
\end{equation}
where the presence of the  index ``r'' identifies the reference-diode quantities; $q_{e}$ is the charge of the electron; and $\mathcal{A}$ is the $^{90}\mathrm{Sr}$ activity, with the factor $2$ to account for the $\upbeta$ electrons emitted from  $^{90}\mathrm{Sr}$ and $^{90}\mathrm{Y}$ decays.

The dependence on the source activity cancels out in the ratio of expected currents,
\begin{equation}
\label{eq:R_exp}
R_{\rm exp} \equiv \frac{I_{\rm exp}(d)}{I_{\rm exp }^{\,\rm r}(d)} = \frac{E (d)}{E_{\rm r}(d) }\frac{E_{eh}^{\, {\rm r}}}{E_{eh}}\,,   
\end{equation}
which is  constant as a function of $d$. This ratio is shown in Figure~\ref{fig:I_vs_d} (b), where it is compared with the ratio  measured from a diamond detector. The expected and measured values are very close to each other.  

\subsubsection{Calibration factors: results}
\label{subsubsec:calibration_results}
We determine the characteristic constant $G$ for each detector from the ratio 
\begin{equation}
\label{eq:G100_def}
\frac{R}{R_{\rm exp}} = \Bigg(\frac{I}{I_{\rm exp}}\Bigg)\,\Bigg(\frac{I^{\,r}_{\rm exp}}{I_{\,r}}\Bigg)  
= \frac{I}{I_{\rm exp}} = G\,,
\end{equation}
where  we assume a precise modelling of the silicon diode ({\it i.e.}  $I_{\rm r} \simeq I_{\rm exp}^{\, \rm r} $).

\begin{figure}[t]
	\centering
	    \begin{overpic}[width=0.9\columnwidth]{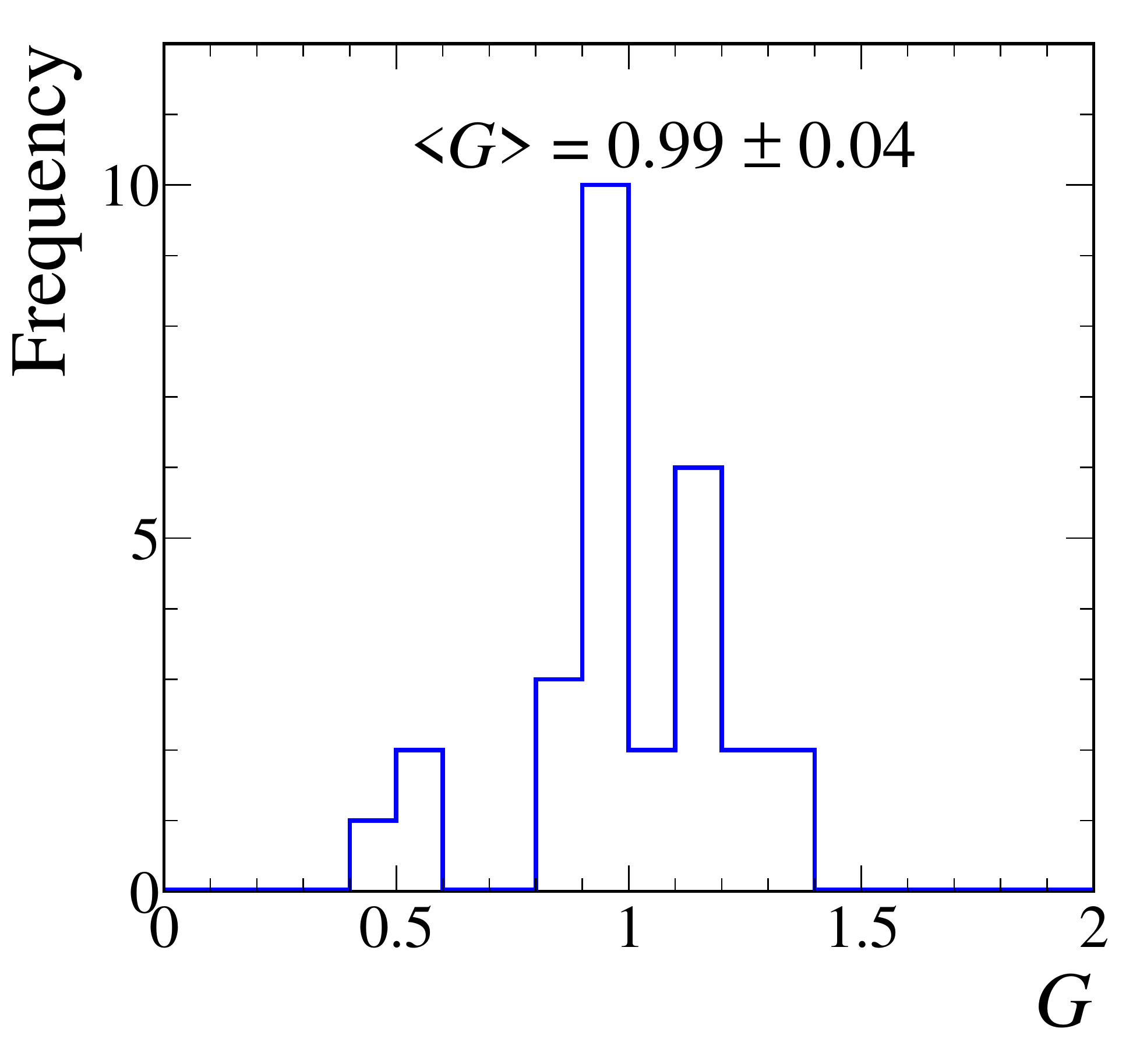}
		\put(20,80){\large (a)}
		\end{overpic}
		\begin{overpic}[width=0.9\columnwidth]{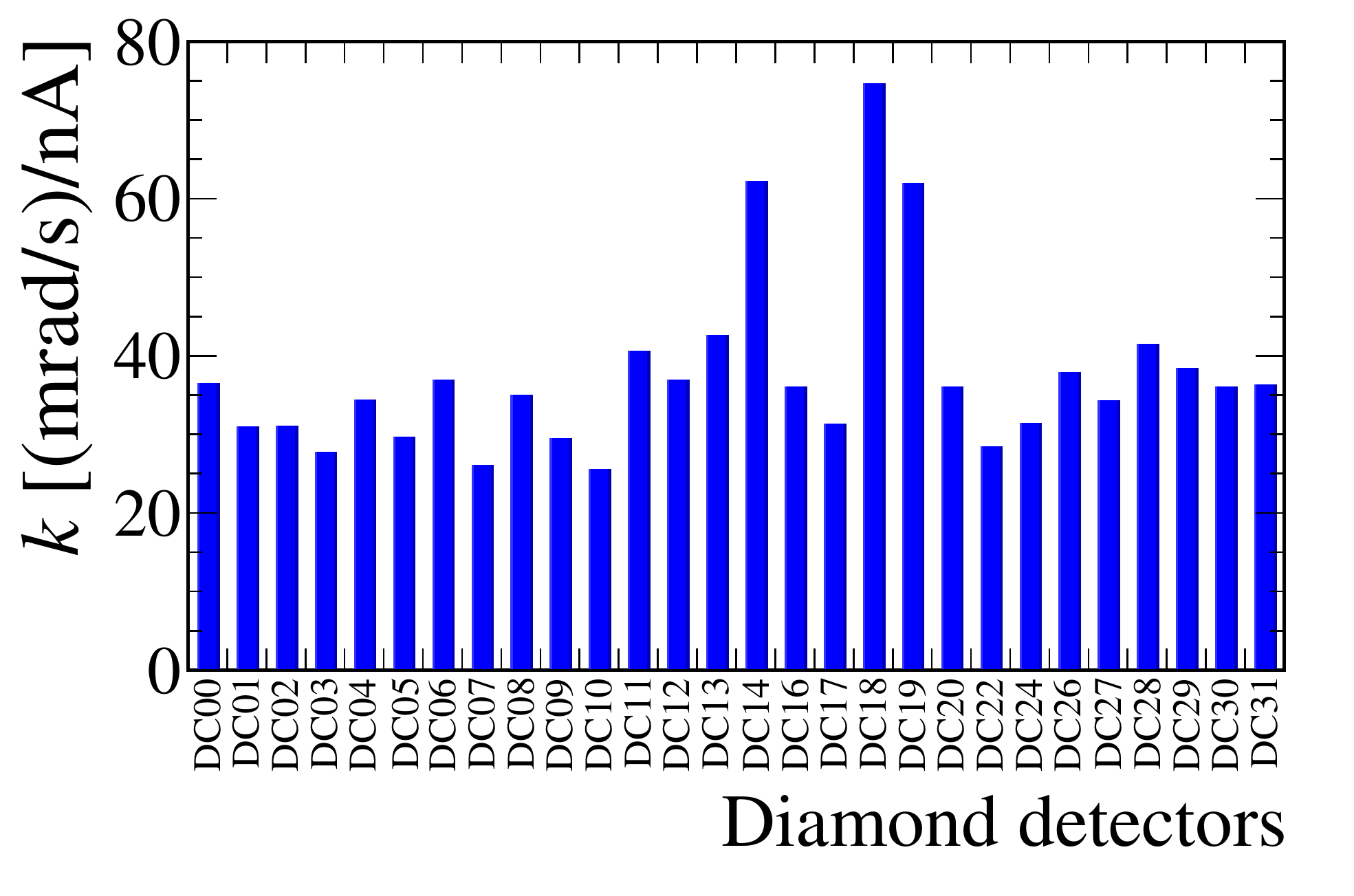}
		\put(20,50){\large (b)}
		\end{overpic}
	\caption{(a) Distribution the of calibration factors $G$ measured from all detectors. (b) The values of the calibration factor $k$ for each detector.}	
	\label{fig:calibration_summary}
\end{figure}

The distribution of $G$ is shown in Figure~\ref{fig:calibration_summary} (a).  The mean value is $\langle G\rangle  = 0.99 \pm 0.04$, which corresponds on average to  fully-efficient detectors, with uniform good-quality crystal ($E_{eh} \approx 13$\,eV) and blocking electrodes (unitary photoconductive gain). Maximum  deviations from this ideal behaviour are limited to about 50\%. 

From the measured $G$ we obtain the  calibration factors reported in Figure~\ref{fig:calibration_summary} (b).
On average, the dose rate is about  $35\,\mathrm{mrad/s}$ for a measured current of  $1$\,nA. The statistical uncertainty on each value of $k$ is negligible with respect to the systematic uncertainty, which is detailed in the next section.

\subsubsection{Systematic uncertainties}
\label{sec:systematics}
The relative uncertainty on each calibration factor totals $\delta k/k=8\%$ from the contributions summarised in Table~\ref{tab:diamonds_gain_systematics} and described as follows. 
\begin{table}[tb]
	\centering
	\resizebox{\columnwidth}{!}{
	\begin{tabular}{ lr }
	\toprule
		\multirow{2}{*}{Systematic-uncertainty source} & $\delta k / k$ \\ 
		  & 	[\%]  \\ 
		\midrule
		Current transients and fluctuations  & $5$ \\
		Measurement of source-detector distance  & $3$ \\
		Silicon ionisation energy & $<1$ \\
		Silicon-diode volume  & $4$ \\
		Package shielding (diode)    & $<1$ \\
		Package shielding (diamond detectors)    & $2$ \\
		Simulation approximations and statistic & $<1$ \\
	    Source description & $<1$ \\
		\midrule
		Total & $8$ \\  
		\bottomrule
	\end{tabular}
	}
	\caption{Relative systematic uncertainties in \% on the calibration factor $k$. }
	\label{tab:diamonds_gain_systematics}
\end{table}

The largest source of uncertainty is due to possible variations of the signal current due to a transient time to reach a stable value. As discussed in Sect.~\ref{subsec:detectors_stability} most  detectors reached a stable output value within a second from the start of irradiation, with small fluctuations below 1\% under steady irradiation. Some other detectors  showed a longer transient time, where the initial values of the current is smaller by approximately 5\% than that  reached after a certain period of steady irradiation. 
To account for the different response of the detectors  to sudden changes of irradiation conditions, which are expected to occur in the SuperKEKB interaction region, we associate a systematic uncertainty of $5\%$ to the measurement of the current, which is propagated to the calibration factor. 

The measured values of $R$ are expected to be constant as a function of $d$, and any trend versus $d$ can be ascribed to a different offset in the source-detector distance between the measurement with the diamond detector and the diode. We observed constant values of $R$ versus $d$ with  a maximum variation of 2\% for some detectors, after correcting the distance with a detector-dependent offset, that averages $0.4$\,mm. Without the offset correction, the deviation from flatness are larger than 2\%, and  the average value of $R$ changes by $3\%$. We assign this change as a systematic uncertainty. 

A major source of uncertainty is related to the reference diode, as we assumed a perfect modelling of its response to obtain $G$. The estimated current from the diode depends on the ionisation energy, on the charge-collection efficiency, and on the energy released in the active volume. 

We propagate to $k$ the uncertainty of $0.03$\,eV on the ionisation energy; this contributes a relative systematic uncertainty of $0.8\%$. 

We assumed a fully efficient diode in collecting all generated charge  at $100$\,V bias.  This assumption is supported by $I$-$V$ measurements under irradiation on the diode, and we consider  the uncertainty on this assumption to be negligible. 

The released energy is proportional to the active volume of the diode used in the simulation, which is known within the uncertainties of the surface and thickness of the detector. To assess a systematic uncertainty on $k$, we make two variations, by changing the active surface and the depletion thickness by $\pm0.2\,\mathrm{mm^2}$ and $\pm 10\,\upmu$m, respectively, which are the associated uncertainties. These variations corresponds to a minimum and a maximum active volume. For both cases, we compute the released energies as a function of the source-detector distance, and the resulting expected ratio $R_{\rm exp}$ from Eq.~\eqref{eq:R_exp}. We take the difference between each new value of $R_{\rm exp}$ and its nominal value, and we consider the largest difference $\delta R_{\rm exp}$ to assign a relative uncertainty $\delta R_{\rm exp}/ R_{\rm exp}$. This is propagated to $\delta k/ k$  and contributes 4\% to the systematic uncertainty.  

In a similar manner, we compute the contribution associated to the uncertainty on the thickness of the aluminium packaging that impacts the released energy. We vary in simulation the thickness of the aluminium cover by $\pm 5\,\upmu$m, which is the uncertainty associated to the cover used in the diode packaging. Other sources that affect the released energy, such as a 0.1\,$\upmu$m layer of silicon oxide on the surface  (the diode area is metallized only around the edge), are considered negligible. 

The value of $R_{\rm exp}$ depends also on the ionisation energy of the diamond detector, on the released energy inside its active volume, and on its charge-collection efficiency. Departures of all these factors from the values assumed in the calculation of the estimated current are accounted for by the characteristic constant $G$. We notice that any variation on $G$ due to an uncertainty on the value of the ionisation energy, or that on the detector volume, are counterbalanced by a variation of $F$ in the definition of $k$ in Eq.~\eqref{eq:k100_def}. Therefore, we do not compute a systematic uncertainty for these two sources.  

The aluminium packaging affects the estimation of the released energy in the diamond detectors in a similar fashion as for the diode, and the impact from the uncertainty on its thickness is computed with the same procedure. We consider an uncertainty of $20\,\upmu$m on the thickness in this case, to account for the different individual covers of the diamond detectors. The resulting systematic uncertainty is $\delta k/k=2\%$. We consider other sources that affect the released energy, {\it e.g.}  the uncertainty on the volume of the electrodes of the diamond detectors, to be negligible. 

As for what concerns the charge-collection efficiency, a departure from the assumed value of 100\% is  included in the characteristic constant $G$ measured for each detector. A possible source of uncertainty is due to small fluctuations of the voltage provided by the power supply, which presented variations of $\delta V/V\approx 0.5\%$. From the $I$-$V$ results presented in Sect.~\ref{subsec:detectors_stability}, we expect only sub-percent changes of the efficiency for $0.5$\,V variation  around a $100$\,V bias. Therefore, we consider negligible the  associated systematic uncertainty on $k$.

The statistical uncertainty from the size of the simulated samples contributes a negligible uncertainty too. Uncertainties related to other assumptions entering the simulation, such as the transport model used in FLUKA for the $\upbeta$ electrons, are found to be smaller than $1\%$, as these contributions are greatly reduced in the ratio of Eq.~\eqref{eq:R_exp}.  For the same reason, an imprecise modelling of the $\upbeta$ source (material, density, fraction of radioactive nuclei) contributes only a sub-$1\%$ systematic uncertainty. 

\subsubsection{Additional checks}
By using the silicon diode as a reference, we do not need  the value of the source activity to determine $G$. Alternatively, the source activity can be measured with the silicon diode through Eq.~\eqref{eq:estimated_current}. In this case, 
we find an activity of about $4.0$\,MBq, to be compared with
a nominal value of approximately $3.0$\,MBq.  We notice that the uncertainty on the nominal activity is not precisely known, and it could be as large as $30\%$.  

If we were to use the nominal activity, $G$ would be computed by the ratio of the current measured from the diamond detector and that computed from Eq.~\eqref{eq:estimated_current}. In this case, the mean values of $G$ would be about $1.5$. Such a value would be explained either by an ionisation energy of about 20\,eV, in disagreement with the TCT measurements reported in Sect.~\ref{subsec:TCT_alpha}, or by a photoconductive gain around $1.5$ for a $100$\,V bias, in disagreement with $I$-$V$ measurements reported in  Sect.~\ref{subsec:detectors_stability}. This would suggest either a large uncertainty on the nominal activity, or a bias on the simulation of the experimental setup, which is greatly reduced in our method by using the diode as a reference through Eq.~\eqref{eq:R_exp}. 

For two detectors, we also carried out two checks of the value of $k$ using photons as a source of radiation. 

In the first check, we exploited soft photons, with a known energy spectrum averaging about 15\,keV, provided by a small X-ray tube~\cite{Gabrielli:2020}. We measured the signal-to-reference ratio of currents at a fixed source-detector distance, using the same silicon diode employed in the measurement with the $\upbeta$ source. The X-ray flux yielded currents between $0.02$\,nA and $0.25$\,nA. We compared the measured ratio with that estimated by means of a full simulation through FLUKA, to obtain $G$ from Eq.~\ref{eq:G100_def}. We found  values of $G$ in agreement with those obtained with the $\upbeta$ radiation. 

With the second check, we aimed at exploring dose rates much higher than those provided by the $\upbeta$ source. We used a source of $\upgamma$ rays from $^{60}$Co decays, with energies of $1.17$\,MeV and $1.33$\,MeV, available  at ISOF-CNR in Bologna.
The source (Gammacell 220) has a certified dosimetry performed with alanine dosimeters in 2007 by Risø National Laboratory. We measured dose rates with the calibrated diamond detectors in
agreement with the certified values, scaled by the  $^{60}$Co reduced activity at the time of the test due to  radiative decay.
We observed a very good linear dependence between the measured currents and the dose rates, spanning a range between $6$\,nA and $130$\,nA  corresponding to dose rates between $0.2\,\mathrm{rad/s}$ and $4.8\,\mathrm{rad/s}$. 
This builds confidence on the validity of the measured calibration factors over a range spanning from tens of $\mathrm{nrad/s}$ to some $\mathrm{rad/s}$.

\section{Conclusions}
\label{conclusions}
We chose  sensors based on single-crystal diamond as dosimeters for a beam-loss monitor at the SuperKEKB electron-positron collider. The diamond crystals, grown by the chemical vapour deposition, are  equipped with Ti+Pt+Au electrodes and assembled in dedicated packages. We presented the test and characterisation of a sample of 28 sensors.

We found that charge-carrier properties and average ionisation energies, as measured by the transient-current technique, are  sufficiently homogeneous across the set of sensors for our purpose.  For about a half of our detectors, we observed  hysteresis effects and unstable currents under constant irradiation  for bias voltages of a specific polarity; stable and reproducible values of the current for the opposite polarity. The other half of the sensors exhibit a symmetric response with respect to bias voltage polarity. We determined the best polarity for each detector, and we found the optimal bias voltage to be $|V|=100$\,V, for which we observed a charge-collection efficiency close to 100\%.

We determined the current-to-dose-rate calibration factors by irradiating the sensors with $\upbeta$ electrons from $^{90}$Sr decays, and by comparing the measured current with that expected from a simulation, used to estimate the dose rate due to the $\upbeta$ irradiation. We employed a silicon diode, irradiated under same conditions, as a reference in order to greatly reduce  uncertainties related to the $\upbeta$-source activity and to the simulation, that would otherwise spoil the accuracy of the calibration. We obtained detector-dependent calibration factors with a relative uncertainty of 8\%. 
We validated the calibration factors with X and $\upgamma$ radiation for dose-rate spanning a range from tens of $\mathrm{nrad/s}$ to some $\mathrm{rad/s}$.

The monitor system based on these detectors has been proving crucial for running the Belle~II experiment in safe conditions during the last two years~\cite{ref:performance}, while beam-background radiation has been continuously evolving with the  progress of the SuperKEKB collider towards unprecedented values of luminosity.  The test and calibration procedures reported here represent valuable resources for the preparation of eight new diamond detectors that will be installed with an upgrade of the Belle~II silicon vertex detector in 2022.

\section*{Acknowledgements}
\label{acknowledgements}
This research was supported by Istituto Nazionale di Fisica Nucleare (INFN) in the framework of the Belle~II experiment. Motivation for improving our calibration procedures was enhanced by useful discussions with SuperKEKB and Belle~II colleagues. We gratefully acknowledge the contribution of Francesco Di Capua (University Federico II and INFN Naples, Italy) in setting up measurements with the $^{60}$Co source.


\bibliography{mybibfile}

\end{document}